\pdfoutput=1

\documentclass[12pt,a4paper]{article}

\usepackage{ifthen} % for conditional statements
\newboolean{pdflatex}
\setboolean{pdflatex}{true} % False for eps figures 

\newboolean{articletitles}
\setboolean{articletitles}{true} % False removes titles in references

\newboolean{uprightparticles}
\setboolean{uprightparticles}{false} %True for upright particle symbols

\newboolean{inbibliography}
\setboolean{inbibliography}{false} %True once you enter the bibliography

\def\paperauthors{LHCb collaboration} % Leave as is for PAPER and CONF
\def\paperasciititle{Search for a dimuon resonance in the Y mass region} % Set ASCII title here
\def\papertitle{Search for a dimuon resonance in the $\Upsilon$ mass region} % Latex formatted title
\def\paperkeywords{{High Energy Physics}, {LHCb}} % Comma separated list
\def\papercopyright{\the\year\ CERN for the benefit of the LHCb collaboration} % new since 9/Apr/2018
\def\paperlicence{CC-BY-4.0 licence}
\def\paperlicenceurl{https://creativecommons.org/licenses/by/4.0/}

\usepackage[top=1in, bottom=1.25in, left=1in, right=1in]{geometry}

\usepackage{microtype}
\usepackage{lineno}  % for line numbering during review
\usepackage{xspace} % To avoid problems with missing or double spaces after
\usepackage{caption} %these three command get the figure and table captions automatically small

\usepackage{graphicx}  % to include figures (can also use other packages)
\usepackage{color}
\usepackage{colortbl}
\graphicspath{{./figs/}} % Make Latex search fig subdir for figures

\usepackage{amsmath} % Adds a large collection of math symbols
\usepackage{amssymb}
\usepackage{amsfonts}
\usepackage{upgreek} % Adds in support for greek letters in roman typeset

\newcommand*\patchAmsMathEnvironmentForLineno[1]{%
\expandafter\let\csname old#1\expandafter\endcsname\csname #1\endcsname
\expandafter\let\csname oldend#1\expandafter\endcsname\csname
end#1\endcsname
 \renewenvironment{#1}%
   {\linenomath\csname old#1\endcsname}%
   {\csname oldend#1\endcsname\endlinenomath}%
}
\newcommand*\patchBothAmsMathEnvironmentsForLineno[1]{%
  \patchAmsMathEnvironmentForLineno{#1}%
  \patchAmsMathEnvironmentForLineno{#1*}%
}
\AtBeginDocument{%
\patchBothAmsMathEnvironmentsForLineno{equation}%
\patchBothAmsMathEnvironmentsForLineno{align}%
\patchBothAmsMathEnvironmentsForLineno{flalign}%
\patchBothAmsMathEnvironmentsForLineno{alignat}%
\patchBothAmsMathEnvironmentsForLineno{gather}%
\patchBothAmsMathEnvironmentsForLineno{multline}%
\patchBothAmsMathEnvironmentsForLineno{eqnarray}%
}

\usepackage{hyperxmp}

\usepackage[pdftex,
            pdfauthor={\paperauthors},
            pdftitle={\paperasciititle},
            pdfkeywords={\paperkeywords},
            pdfcopyright={Copyright (C) \papercopyright},
            pdflicenseurl={\paperlicenceurl}]{hyperref}

\usepackage[all]{hypcap} % Internal hyperlinks to floats.

\usepackage{xspace} 
\usepackage{upgreek}

\newcommand{\Azo}{\ensuremath{\phi}\xspace}

\newcommand{\AMuMu}{\ensuremath{\Azo\to\mu^+\mu^-}\xspace}

\def\UpsilonNS{\ensuremath{\Upsilon{(nS)}}\xspace}% no space before {...}!
\def\UpsilonOS{\ensuremath{\Upsilon{(1S)}}\xspace}% no space before {...}!
\newcommand{\TeV}{\ensuremath{\,{\rm TeV}}\xspace}

\def\lhcb {\mbox{LHCb}\xspace}

\def\MagUp {\mbox{\em Mag\kern -0.05em Up}\xspace}

\ifthenelse{\boolean{uprightparticles}}%
{

 \def\Pmu         {\ensuremath{\upmu}\xspace}

 \def\Ppsi        {\ensuremath{\uppsi}\xspace}

 \def\PDelta      {\ensuremath{\Delta}\xspace}                 
 \def\PXi      {\ensuremath{\Xi}\xspace}                 
 \def\PLambda      {\ensuremath{\Lambda}\xspace}                 
 \def\PSigma      {\ensuremath{\Sigma}\xspace}                 
 \def\POmega      {\ensuremath{\Omega}\xspace}                 
 \def\PUpsilon      {\ensuremath{\Upsilon}\xspace}                 
 
 \def\PB      {\ensuremath{\mathrm{B}}\xspace}                 
                  
 \def\PD      {\ensuremath{\mathrm{D}}\xspace}

 \def\PJ      {\ensuremath{\mathrm{J}}\xspace}                 
 \def\PK      {\ensuremath{\mathrm{K}}\xspace}

 \def\Pb      {\ensuremath{\mathrm{b}}\xspace}                 
 \def\Pc      {\ensuremath{\mathrm{c}}\xspace}

 \def\Pi      {\ensuremath{\mathrm{i}}\xspace}

}
{

 \def\Pmu         {\ensuremath{\mu}\xspace}

 \def\Ppsi        {\ensuremath{\psi}\xspace}                 
                  
 \mathchardef\PDelta="7101
 \mathchardef\PXi="7104
 \mathchardef\PLambda="7103
 \mathchardef\PSigma="7106
 \mathchardef\POmega="710A
 \mathchardef\PUpsilon="7107
                  
 \def\PB      {\ensuremath{B}\xspace}                 
                  
 \def\PD      {\ensuremath{D}\xspace}

 \def\PJ      {\ensuremath{J}\xspace}                 
 \def\PK      {\ensuremath{K}\xspace}

 \def\Pb      {\ensuremath{b}\xspace}                 
 \def\Pc      {\ensuremath{c}\xspace}

 \def\Pi      {\ensuremath{i}\xspace}

}

\makeatletter
\ifcase \@ptsize \relax% 10pt
  \newcommand{\miniscule}{\@setfontsize\miniscule{4}{5}}% \tiny: 5/6
\or% 11pt
  \newcommand{\miniscule}{\@setfontsize\miniscule{5}{6}}% \tiny: 6/7
\or% 12pt
  \newcommand{\miniscule}{\@setfontsize\miniscule{5}{6}}% \tiny: 6/7
\fi
\makeatother

\DeclareRobustCommand{\optbar}[1]{\shortstack{{\miniscule (\rule[.5ex]{1.25em}{.18mm})}
  \\ [-.7ex] $#1$}}

\def\mup        {{\ensuremath{\Pmu^+}}\xspace}
\def\mun        {{\ensuremath{\Pmu^-}}\xspace} % muon negative (\mum is taken)
\def\mumu       {{\ensuremath{\Pmu^+\Pmu^-}}\xspace}

\def\cquark    {{\ensuremath{\Pc}}\xspace}

\def\bquark    {{\ensuremath{\Pb}}\xspace}

  \def\Kbar    {{\kern 0.2em\overline{\kern -0.2em \PK}{}}\xspace}

\def\KorKbar    {\kern 0.18em\optbar{\kern -0.18em K}{}\xspace}

  \def\Dbar    {{\kern 0.2em\overline{\kern -0.2em \PD}{}}\xspace}

\def\DorDbar    {\kern 0.18em\optbar{\kern -0.18em D}{}\xspace}

\def\Bbar    {{\ensuremath{\kern 0.18em\overline{\kern -0.18em \PB}{}}}\xspace}

\def\BorBbar    {\kern 0.18em\optbar{\kern -0.18em B}{}\xspace}

\def\jpsi     {{\ensuremath{{\PJ\mskip -3mu/\mskip -2mu\Ppsi\mskip 2mu}}}\xspace}

  \def\Y#1S{\ensuremath{\PUpsilon{(#1S)}}\xspace}% no space before {...}!

\def\Lbar        {{\ensuremath{\kern 0.1em\overline{\kern -0.1em\PLambda}}}\xspace}
\def\LorLbar    {\kern 0.18em\optbar{\kern -0.18em \PLambda}{}\xspace}

\def\BF         {{\ensuremath{\mathcal{B}}}\xspace}

\def\to                 {\ensuremath{\rightarrow}\xspace}

\def\Azero         {\Azo}

\newcommand{\BR}[1]{\ensuremath{{\cal B}(#1)}\xspace}

\def\AT#1     {\ensuremath{A_{\mathrm{T}}^{#1}}\xspace}           % 2

\def\C#1      {\ensuremath{\mathcal{C}_{#1}}\xspace}                       % 9
\def\Cp#1     {\ensuremath{\mathcal{C}_{#1}^{'}}\xspace}                    % 7
\def\Ceff#1   {\ensuremath{\mathcal{C}_{#1}^{\mathrm{(eff)}}}\xspace}        % 9  
\def\Cpeff#1  {\ensuremath{\mathcal{C}_{#1}^{'\mathrm{(eff)}}}\xspace}       % 7
\def\Ope#1    {\ensuremath{\mathcal{O}_{#1}}\xspace}                       % 2
\def\Opep#1   {\ensuremath{\mathcal{O}_{#1}^{'}}\xspace}                    % 7

\newcommand{\tev}{\ifthenelse{\boolean{inbibliography}}{\ensuremath{~T\kern -0.05em eV}}{\ensuremath{\mathrm{\,Te\kern -0.1em V}}}\xspace}
\newcommand{\gev}{\ensuremath{\mathrm{\,Ge\kern -0.1em V}}\xspace}

\newcommand{\mev}{\ensuremath{\mathrm{\,Me\kern -0.1em V}}\xspace}
\newcommand{\kev}{\ensuremath{\mathrm{\,ke\kern -0.1em V}}\xspace}
\newcommand{\ev}{\ensuremath{\mathrm{\,e\kern -0.1em V}}\xspace}
\newcommand{\gevc}{\ensuremath{{\mathrm{\,Ge\kern -0.1em V\!/}c}}\xspace}
\newcommand{\mevc}{\ensuremath{{\mathrm{\,Me\kern -0.1em V\!/}c}}\xspace}
\newcommand{\gevcc}{\ensuremath{{\mathrm{\,Ge\kern -0.1em V\!/}c^2}}\xspace}
\newcommand{\gevgevcccc}{\ensuremath{{\mathrm{\,Ge\kern -0.1em V^2\!/}c^4}}\xspace}
\newcommand{\mevcc}{\ensuremath{{\mathrm{\,Me\kern -0.1em V\!/}c^2}}\xspace}

\def\mum  {\ensuremath{{\,\upmu\mathrm{m}}}\xspace}

\def\pb {\ensuremath{\mathrm{ \,pb}}\xspace}

\def\invfb   {\ensuremath{\mbox{\,fb}^{-1}}\xspace}

\def\ps   {\ensuremath{{\mathrm{ \,ps}}}\xspace}

\newcommand{\chisq}{\ensuremath{\chi^2}\xspace}

\def\gsim{{~\raise.15em\hbox{$>$}\kern-.85em
          \lower.35em\hbox{$\sim$}~}\xspace}
\def\lsim{{~\raise.15em\hbox{$<$}\kern-.85em
          \lower.35em\hbox{$\sim$}~}\xspace}

\def\sPlot{\mbox{\em sPlot}\xspace}
\def\ptot       {\mbox{$p$}\xspace}
\def\pt         {\mbox{$p_{\mathrm{ T}}$}\xspace}

\def\evtgen     {\mbox{\textsc{EvtGen}}\xspace}

\def\geant      {\mbox{\textsc{Geant4}}\xspace}

\def\photos     {\mbox{\textsc{Photos}}\xspace}

\def\pythia     {\mbox{\textsc{Pythia}}\xspace}

\def\tell1  {TELL1\xspace}
\def\ukl1   {UKL1\xspace}

\newcommand{\eg}{\mbox{\itshape e.g.}\xspace}

\usepackage{cite} % Allows for ranges in citations
\usepackage{mciteplus}

\newcommand{\figref}[1]{Fig.~\ref{#1}}

\newcommand{\appref}[1]{Appendix~\ref{#1}}

\usepackage{longtable} % only for template; not usually to be used in PAPERs

\begin{document}

%%%%%%%%%%%%%%%%%%%%%%%%%
%%%%% Title     %%%%%%%%%
%%%%%%%%%%%%%%%%%%%%%%%%%
\renewcommand{\thefootnote}{\fnsymbol{footnote}}
\setcounter{footnote}{1}

% %%%%%%% CHOOSE TITLE PAGE--------
%\onecolumn
%\input{title-LHCb-INT}
%\input{title-LHCb-ANA}
%\input{title-LHCb-CONF}
% $Id: title-LHCb-PAPER.tex 118821 2018-04-09 19:15:15Z pkoppenb $
% ===============================================================================
% Purpose: LHCb-PAPER journal paper title page template
% Author: 
% Created on: 2010-09-25
% ===============================================================================

%%%%%%%%%%%%%%%%%%%%%%%%%
%%%%%  TITLE PAGE  %%%%%%
%%%%%%%%%%%%%%%%%%%%%%%%%
\begin{titlepage}
\pagenumbering{roman}

% Header ---------------------------------------------------
\vspace*{-1.5cm}
\centerline{\large EUROPEAN ORGANIZATION FOR NUCLEAR RESEARCH (CERN)}
\vspace*{1.5cm}
\noindent
\begin{tabular*}{\linewidth}{lc@{\extracolsep{\fill}}r@{\extracolsep{0pt}}}
\ifthenelse{\boolean{pdflatex}}% Logo format choice
{\vspace*{-1.5cm}\mbox{\!\!\!\includegraphics[width=.14\textwidth]{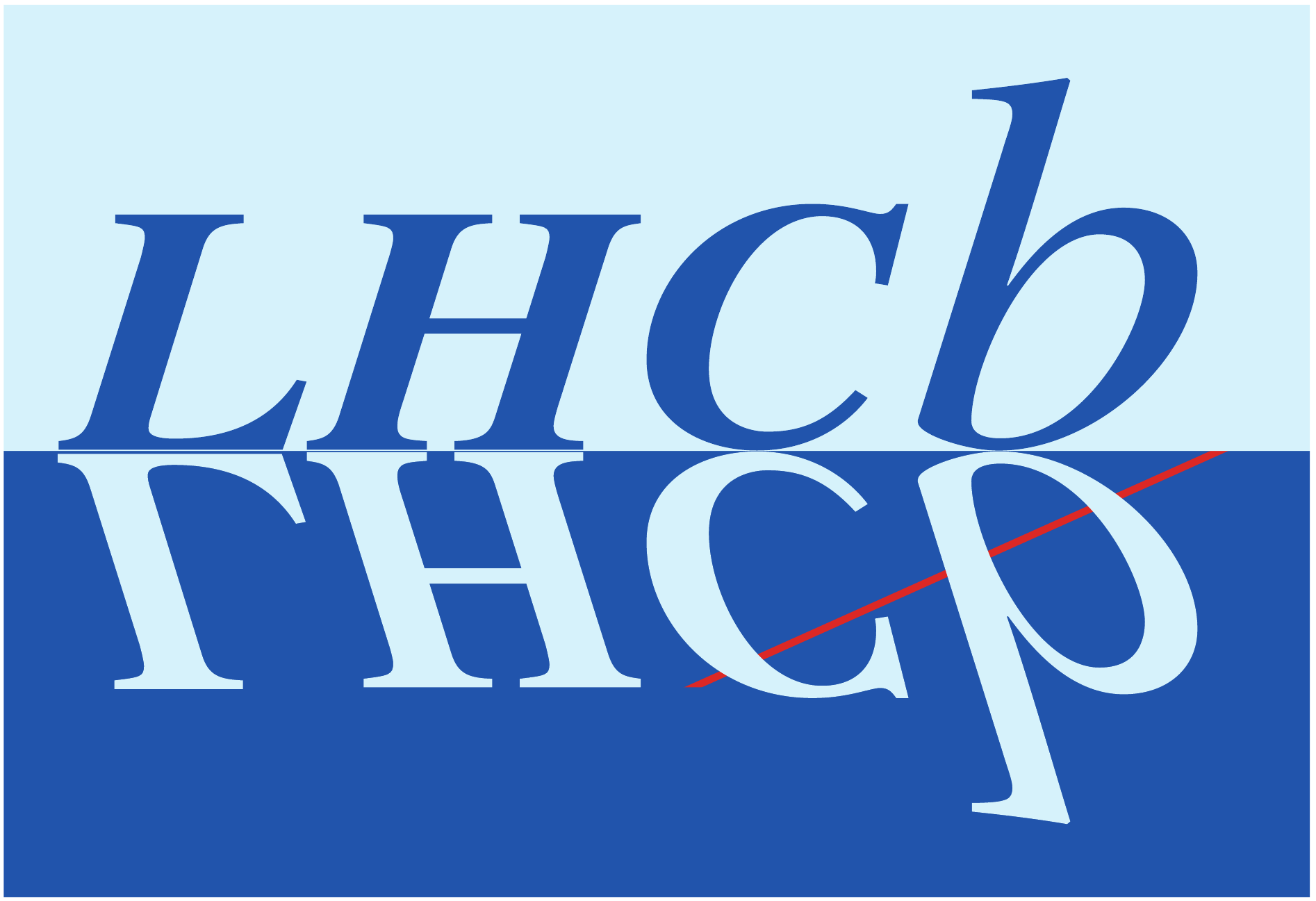}} & &}%
{\vspace*{-1.2cm}\mbox{\!\!\!\includegraphics[width=.12\textwidth]{lhcb-logo.eps}} & &}%
\\
 & & CERN-EP-2018-111 \\  % ID 
 & & LHCb-PAPER-2018-008 \\  % ID 
 & & 27 Sep 2018 \\ % Date - Can also hardwire e.g.: 23 March 2010
 & & \\
% not in paper \hline
\end{tabular*}

\vspace*{4.0cm}

% Title --------------------------------------------------
{\normalfont\bfseries\boldmath\huge
\begin{center}
% DO NOT EDIT HERE. Instead edit macro in main.tex to keep metadata correct
  \papertitle 
\end{center}
}

\vspace*{2.0cm}

% Authors -------------------------------------------------
\begin{center}
%In the footnote, replace 'paper' by 'Letter' in case of submission to PRL or PLB 
% Edit macro in main.tex to keep metadata correct
\paperauthors\footnote{Authors are listed at the end of this paper.}
\end{center}

\vspace{\fill}

% Abstract -----------------------------------------------
\begin{abstract}
  \noindent A search is performed for a spin-0 boson, $\phi$, produced in proton-proton collisions at centre-of-mass energies of 7 and 8\tev, using prompt $\Azo\to\mumu$ decays and a data sample corresponding to an integrated luminosity of approximately 3.0\invfb collected with the LHCb detector.
  No evidence is found for a signal in the mass range from $5.5$ to $15\gev$. Upper limits are placed on the product of the production cross-section and the branching fraction into the dimuon final state.
  The limits are comparable to the best existing over most of the mass region considered and are the first to be set near the $\Upsilon$ resonances. 
\end{abstract}

\vspace*{2.0cm}

\begin{center}
  Published in JHEP
\end{center}

\vspace{\fill}

{\footnotesize 
% Edit macro in main.tex to keep metadata correct
\centerline{\copyright~\papercopyright. \href{\paperlicenceurl}{\paperlicence}.}}
\vspace*{2mm}

\end{titlepage}

%%%%%%%%%%%%%%%%%%%%%%%%%%%%%%%%
%%%%%  EOD OF TITLE PAGE  %%%%%%
%%%%%%%%%%%%%%%%%%%%%%%%%%%%%%%%

%  empty page follows the title page ----
\newpage
\setcounter{page}{2}
\mbox{~}
%\newpage
%
%% Author List ----------------------------
%%  You need to get a new author list!
%\input{LHCb_authorlist.tex}
%
%The author list for journal publications is provided by the Membership Committee shortly after 'approval to go to paper' has been given.
%%It will be made available on the page
%%\verb!http://www.physik.uzh.ch/~strauman/forMemCo/LHCb-PAPER-XXXX-XXX/! .
%It will be sent to you by email shortly after a paper number has beens assigned.
%The author list should be included already at first circulation, 
%to allow new members of the collaboration to verify whether they have been included correctly.
%Occasionally a misspelled name is corrected or associated institutions become full members.
%In that case, a new author list will be sent to you.
%In case line numbering doesn't work well after including the authorlist, try moving the \verb!\bigskip! after the last author to a separate line.
%
%
%The authorship for Conference Reports should be ``The LHCb
%  collaboration'', with a footnote giving the name(s) of the contact
%  author(s), but without the full list of collaboration names.

\cleardoublepage

%\twocolumn
% %%%%%%%%%%%%% ---------

\renewcommand{\thefootnote}{\arabic{footnote}}
\setcounter{footnote}{0}

%%%%%%%%%%%%%%%%%%%%%%%%%%%%%%%%
%%%%%  Table of Content   %%%%%%
%%%%%%%%%%%%%%%%%%%%%%%%%%%%%%%%
%%%% Uncomment next 2 lines if desired
%\tableofcontents
%\cleardoublepage

%%%%%%%%%%%%%%%%%%%%%%%%%
%%%%% Main text %%%%%%%%%
%%%%%%%%%%%%%%%%%%%%%%%%%

\pagestyle{plain} % restore page numbers for the main text
\setcounter{page}{1}
\pagenumbering{arabic}

%% Uncomment during review phase. 
%% Comment before a final submission.
%\linenumbers

% You can include short sections directly in the main tex file.
% However, for larger papers it is desirable to split the text into
% several semiautonomous files, which can be revised independently.
% This is especially useful when developing a document in
% collaboration with several people, since then different parts can be
% edited independently.  This type of file organization is shown here.
% 

\section{Introduction}
\label{sec:Introduction}

The only known elementary spin-0 particle is the resonance of mass 125 GeV ($c=1$ throughout this paper) discovered at the LHC, $H$, whose properties are found to be consistent with those of the Standard Model (SM) Higgs boson~\cite{Chatrchyan:2012xdj, Aad:2012tfa}. 
However, additional spin-0 bosons, $\Azo$, arise in many extensions of the SM and are often predicted to be lighter than the $H$ boson mass, $m(H)$~\cite{Curtin:2013fra}. Examples of models with light (pseudo-)scalar particles are the next-to-minimal supersymmetric SM (NMSSM)~\cite{Fayet:1976cr, PhysRevD.39.844, Ellwanger:2009dp}, Little Higgs models~\cite{ArkaniHamed:2001nc, ArkaniHamed:2002qx, Perelstein:2005ka} and the two-Higgs
doublet model with an additional scalar~\cite{Curtin:2013fra}. 
Scalar fields can also provide portals to so-called dark sectors that are neutral under SM interactions and that might include dark matter particles~\cite{darksector1, darksector2, darksector3}. A scalar portal mediated by a light particle can also be associated to the inflation of the early Universe~\cite{Bezrukov:2009yw, Bezrukov:2013fca}.

An extensive and diverse set of searches has been performed for new spin-0 particles with masses less than $m(H)$ (see Ref.~\cite{Haisch:2018kqx} for a recent review). Most searches performed by the ATLAS and CMS collaborations rely on the hypothetical decay $H\to\Azo\Azo$ and on the reconstruction of the two $\Azo$ boson decays in the $\mu^+\mu^-$, $\tau^+\tau^-$ and $b\bar{b}$ final states. 
A complementary strategy~\cite{Haisch:2018kqx} consists of searching for the direct production of $\Azo$ bosons in $pp$ collisions via, \eg gluon-gluon fusion. Searches of this type performed at the LHC have aimed at reconstructing a possible $\Azo$ boson in its decay to either $\gamma\gamma$, $\tau^+\tau^-$ or $\mumu$. A recent search in the $\gamma\gamma$ final state~\cite{CMS:2017yta} explored a mass range down to $m(\Azo)=70\gev$, while one employing $\tau^+\tau^-$ explored masses down to
$m(\Azo)=90\gev$~\cite{Sirunyan:2018zut}. Masses as low as $m(\Azo)=25\gev$ were also investigated in the $\Azo\to\tau^+\tau^-$ decay using the signature of a $\Azo$ boson produced in association with two $b$ jets~\cite{Khachatryan:2015baw}. 
For lower masses, searches in the dimuon spectrum are currently the most sensitive~\cite{Haisch:2018kqx} and include \Azo bosons produced in either gluon-gluon fusion in LHC collisions~\cite{Chatrchyan:2012am}, \UpsilonOS radiative decays~\cite{Lees:2012iw, Lees:2012te} or rare $b$-hadron decays~\cite{LHCb-PAPER-2015-036,LHCb-PAPER-2016-052}.

As shown in Ref.~\cite{Haisch:2016hzu}, the LHCb detector has good sensitivity to light spin-0 particles due to its high-precision spectrometer and its capability of triggering on objects with small transverse momenta.
LHCb has already searched for prompt dark photons decaying to dimuons with invariant masses up to 70\gev~\cite{LHCb-PAPER-2017-038} using $pp$ collisions at 13\tev corresponding to an integrated luminosity of 1.6\invfb. These results were recently reinterpreted in the context of a $\Azo$ boson search and provide the best limits in the mass region between 10.6 to 70 \gev~\cite{Haisch:2018kqx}, even though this search was optimised for the dark photon production kinematics. However, all
searches in $pp$ collisions exclude the region dominated by $\Upsilon$ resonances. 
 
This article presents a search for a narrow dimuon resonance in the mass region between 5.5 and 15\gev. The excellent mass resolution of the \lhcb detector is exploited to study the region close to the $\Upsilon$ resonances that was not explored in previous searches. For this analysis, signal candidates are selected from $pp$ collision data corresponding to an integrated luminosity of $0.98~(1.99) \invfb$, recorded with the LHCb detector during 2011 (2012) at a centre-of-mass energy of $\sqrt{s}=$7 (8) $\TeV$ (a data set statistically independent from that of Ref.~\cite{LHCb-PAPER-2017-038}). 

The results are interpreted in the context of a \Azo boson produced directly in the $pp$ collision through gluon-gluon fusion. The analysis has been designed in a model-independent way for any prompt dimuon resonance, be it predicted by the SM (\eg $\eta_b\to\mumu$ as suggested in Ref.~\cite{Haisch:2016hzu}) or not. 
In order to be independent of the production mechanism, the data set is analysed separately in bins of the dimuon kinematics and for the two collision energies. The results are also independent of the resonance spin, allowing for an interpretation in terms of a vector boson, $A^\prime$.

\section{Detector and simulation}
\label{sec:detector}

The \lhcb detector~\cite{Alves:2008zz,LHCb-DP-2014-002} is a single-arm forward
spectrometer covering the pseudorapidity range $2<\eta <5$,
designed for the study of particles containing \bquark or \cquark
quarks. The detector includes a high-precision tracking system
consisting of a silicon-strip vertex detector surrounding the $pp$
interaction region~\cite{LHCb-DP-2014-001}, a large-area silicon-strip detector located
upstream of a dipole magnet with a bending power of about
$4{\mathrm{\,Tm}}$, and three stations of silicon-strip detectors and straw
drift tubes~\cite{LHCb-DP-2013-003} placed downstream of the magnet.
The tracking system provides a measurement of momentum, \ptot, of charged particles with
a relative uncertainty that varies from 0.5\% at low momentum to 1.0\% at 200\gev.
The minimum distance of a track to a primary vertex (PV), the impact parameter (IP), 
is measured with a resolution of $(15+29\gev/\pt)\mum$,
where \pt is the component of the momentum transverse to the beam.
Different types of charged hadrons are distinguished using information from 
two ring-imaging Cherenkov detectors~\cite{LHCb-DP-2012-003}. 
Photons, electrons and hadrons are identified by a calorimeter system consisting of
scintillating-pad (SPD) and preshower detectors, an electromagnetic
calorimeter and a hadronic calorimeter.
Muons are identified by a
system composed of alternating layers of iron and multiwire
proportional chambers~\cite{LHCb-DP-2012-002}. 

The online event selection is performed by a trigger~\cite{LHCb-DP-2012-004}, 
which consists of a hardware stage, based on information from the calorimeter and muon
systems, followed by a software stage, which applies a full event
reconstruction. In this analysis,  signal candidates are first required
to pass the hardware trigger, which selects events containing at least
one muon with $\pt>1.5$ $(1.8)\gev$ in the 7 (8)\tev data sample.  The subsequent software
trigger requires events with either a muon with $\pt>10\gev$, or alternatively, a 
pair of muons having an invariant mass larger than $4.7\gev$, forming a good quality vertex 
and with the larger muon \pt exceeding $4.8\gev$.      
A global event cut (GEC) is also applied at the hardware stage, which requires the number 
of hits in the SPD to be less than 600. 
 
In the simulation, $pp$ collisions are generated using
\pythia~\cite{Sjostrand:2007gs,Sjostrand:2006za} 
 with a specific \lhcb configuration~\cite{LHCb-PROC-2010-056}.  Decays of hadronic particles
are described by \evtgen~\cite{Lange:2001uf}, in which final-state
radiation is generated using \photos~\cite{Golonka:2005pn}. The
interaction of the generated particles with the detector, and its response,
are implemented using the \geant
toolkit~\cite{Allison:2006ve, *Agostinelli:2002hh} as described in
Ref.~\cite{LHCb-PROC-2011-006}.

\section{Event selection}
\label{sec:selection}
A dimuon candidate is formed using two oppositely charged tracks identified as muons, which must satisfy the requirements of the hardware and software stages of the trigger.
The vertex formed by the two tracks is required to be of good quality and to be consistent with the location of the primary vertex. 
Finally, the reconstructed proper decay time is required to be less than $0.1\ps$ to suppress background from muons produced in the decays of heavy flavour hadrons.

The dimuon invariant mass spectrum is investigated in the range from $5.5\gev$, above the region dominated by $b$--hadron decays, up to $15\gev$. In this mass region the $m(\Azo)$ resolution is about $0.5\%$ and the total acceptance for \Azo bosons produced via gluon-gluon fusion is between 2 and 3\%.

\begin{figure}[t]
\begin{center}
\includegraphics[width=0.68\textwidth]{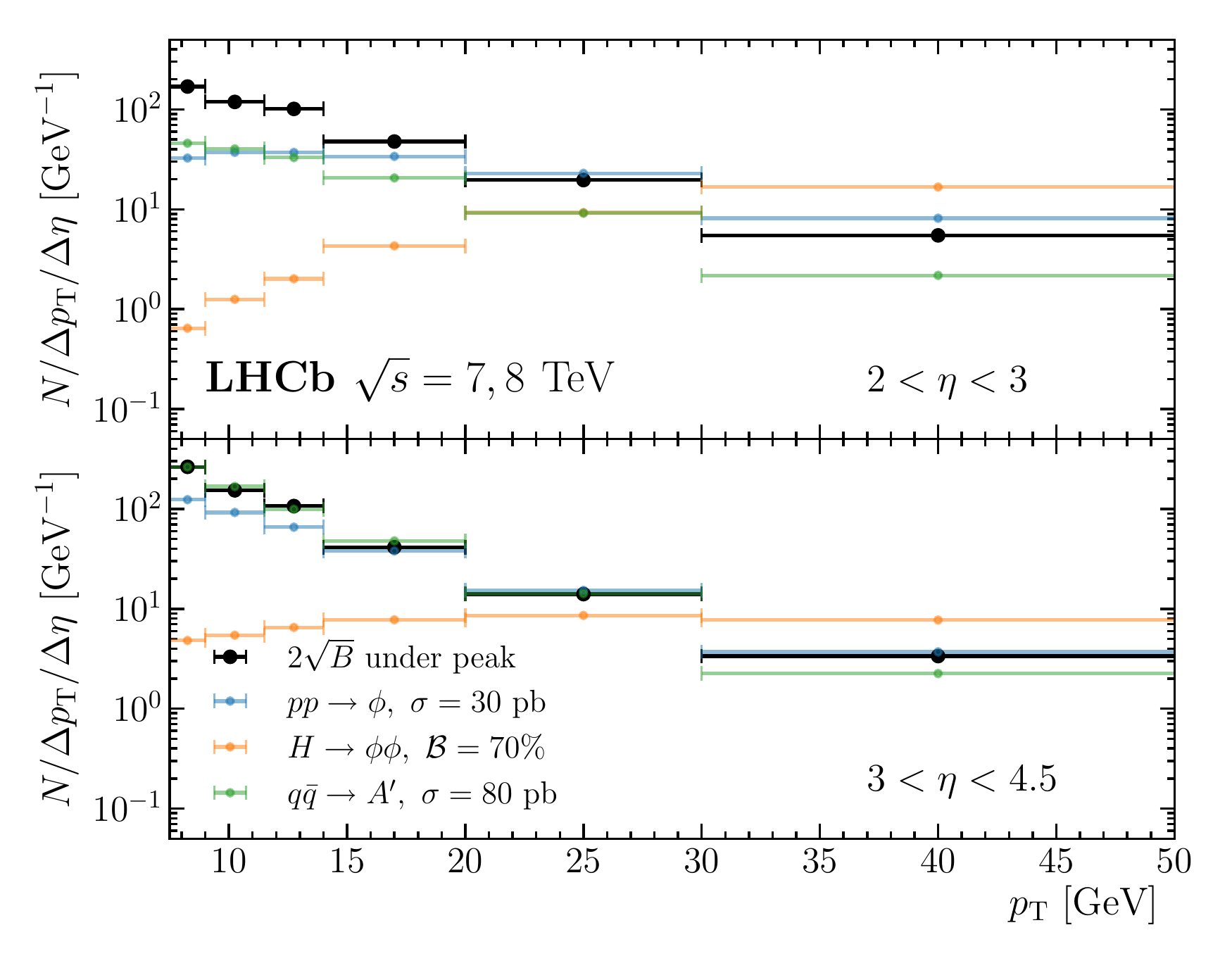}
\caption{The expected sensitivity, defined as $2\sqrt{B}$, where $B$ is the background under a dimuon peak with invariant mass ${11\gev}$, is shown for the 12 $[\pt, \eta]$ bins. For comparison, signal yields in the various bins are shown for three different production mechanisms: a \Azo boson produced via gluon-gluon fusion, a \Azo boson coming from a $H\to\Azo\Azo$ decay and a vector $A^\prime$ boson produced via the Drell-Yan mechanism.}
\label{fig:sig_fractions}
\end{center}
\end{figure}

A fiducial region is defined for the kinematics of the dimuon candidate: each muon is required to be within $2.0<\eta<4.9$, and the higher (lower) muon \pt is required to be in excess of  $4.8\gev$ ($2.5\gev$). Moreover, the $\Azo$ boson candidate \pt is required to be between 7.5 and 50\gev and its pseudorapidity between 2 and 4.5. The search is then performed in 6 bins of $\pt(\Azo)$ and 2 bins of $\eta(\Azo)$ as well as separately for the two $pp$ collision energies, for a total of 24 separate samples. 
As shown in  \figref{fig:sig_fractions}, the binned analysis provides better separation of signal from background if the \Azo boson production spectrum is significantly different from that of the background dimuon candidates.
 
%
%%%%%%%

Apart from the narrow \UpsilonNS ($n=1,2,3$) resonances, the selected candidates are composed of three categories: genuine muon pairs produced via the Drell-Yan mechanism, pairs of displaced muons coming from heavy flavour decays, and wrong associations of one such muon with a prompt pion that is misidentified as a muon. While the Drell-Yan component is indistinguishable from a signal with the same production spectrum, the other two categories can be reduced. 
For this purpose, a multivariate (MVA) classifier based on the uniform boosting (uBoost) algorithm~\cite{uBoost} is used, where a boosted decision tree is trained to separate signal from background candidates. 
This technique has been successfully used in previous LHCb searches~\cite{LHCb-PAPER-2015-036}, as it avoids biasing the mass spectrum and, most importantly, it simplifies the determination of the classification efficiency, which can be evaluated for a single mass using, for example, \UpsilonOS data.  
%%%%%%%
The MVA classifier is trained using a signal sample consisting of simulated Drell-Yan events and a background data sample composed of pairs of muon candidates with the same electric charge. 

The classifier is trained on the following kinematic and topological features: IP, \pt, momentum and track-fit \chisq of each muon candidate; minimum IP \chisq of both muons with respect to any PV in the event, where the IP \chisq is defined as the difference between the vertex-fit \chisq of a PV reconstructed with and without the muon; the angle between the positive muon in the \Azo boson rest frame and the direction opposite to that of the \Azo boson in the laboratory frame; IP of the dimuon
candidate; and the isolation variable defined in Ref.~\cite{LHCb-PAPER-2013-046}, related to the number of good two-track vertices a muon can make with other tracks in the event, to reduce the background from heavy flavour decays. 

In order to account for small differences between simulation and data, a correction is applied through a multi-dimensional weighting \cite{hep_ml}. This correction is determined by matching simulation and data in various detector-related variables of a \UpsilonOS sample. Examples of the variables showing discrepancies are the track-fit \chisq and the IP \chisq of the muons. For the data sample, background is statistically subtracted using the \sPlot technique~\cite{Pivk:2004ty} based
on a fit to the \UpsilonOS dimuon mass peak. 

To determine the best MVA requirement, the ratio ${\epsilon_S / (3/2 + \sqrt{B} )}$~\cite{Punzi:2003bu} is maximised, where $\epsilon_S$ is the signal efficiency and $B$ the mean background yield. For this, $\epsilon_S$ is taken from $pp\to \AMuMu$ simulated samples, while an estimate of the average background yield under the hypothetical signal peak is taken from the mass sidebands
of the \UpsilonNS region in data.  
The resulting MVA requirement is about $90\%$ efficient on reconstructed $pp\to \AMuMu$ signal while it reduces the background by about $40\%$. By comparing the samples composed of same-sign and opposite-sign muons, the genuine dimuon purity is estimated to be about $50\%$.

%%%%%%%

\section{Signal efficiencies}
\label{sec:efficiencies}
The determination of signal efficiencies relies on simulated dimuon samples, which are corrected for small inaccuracies of the simulation using control data samples. 

Trigger efficiencies are above $90\%$. They are determined from simulation and checked on data. The efficiency of the global event cut is instead taken from data using a sample of \UpsilonOS candidates selected in the hardware trigger using a much looser requirement on the SPD multiplicity.
Given the event multiplicity does not significantly change with dimuon mass, the same GEC efficiency is used for the whole range of masses.

The reconstruction and selection efficiencies are determined using simulation. The muon track reconstruction efficiency is corrected as a function of the track kinematics using a data sample of $J/\psi\to\mu^+\mu^-$ decays~\cite{LHCb-DP-2013-002}. The total systematic uncertainty related to this procedure is about $0.8\%$.

The efficiency of the MVA selection is computed using the weighted simulation sample and is tested on \UpsilonOS candidates selected without applying the MVA criterion. 
The efficiency difference in each bin is below $2\%$, which is assigned as a systematic uncertainty. 
The MVA response is decorrelated from the dimuon mass due to the use of the uBoost technique allowing this cross-check to be valid for the whole range of $m(\Azo)$ considered. 

The muon identification efficiency is calculated using a sample of $\jpsi\to\mumu$ decays, following the procedure in Ref.~\cite{LHCb-PUB-2016-021}. 
In addition to the statistical uncertainty due to the finite size of the calibration sample, a systematic uncertainty between $1$ and $7\%$ is assigned due to the finite width of the kinematic bins used.

\begin{figure}[t]
\begin{center}
\includegraphics[width=0.68\textwidth]{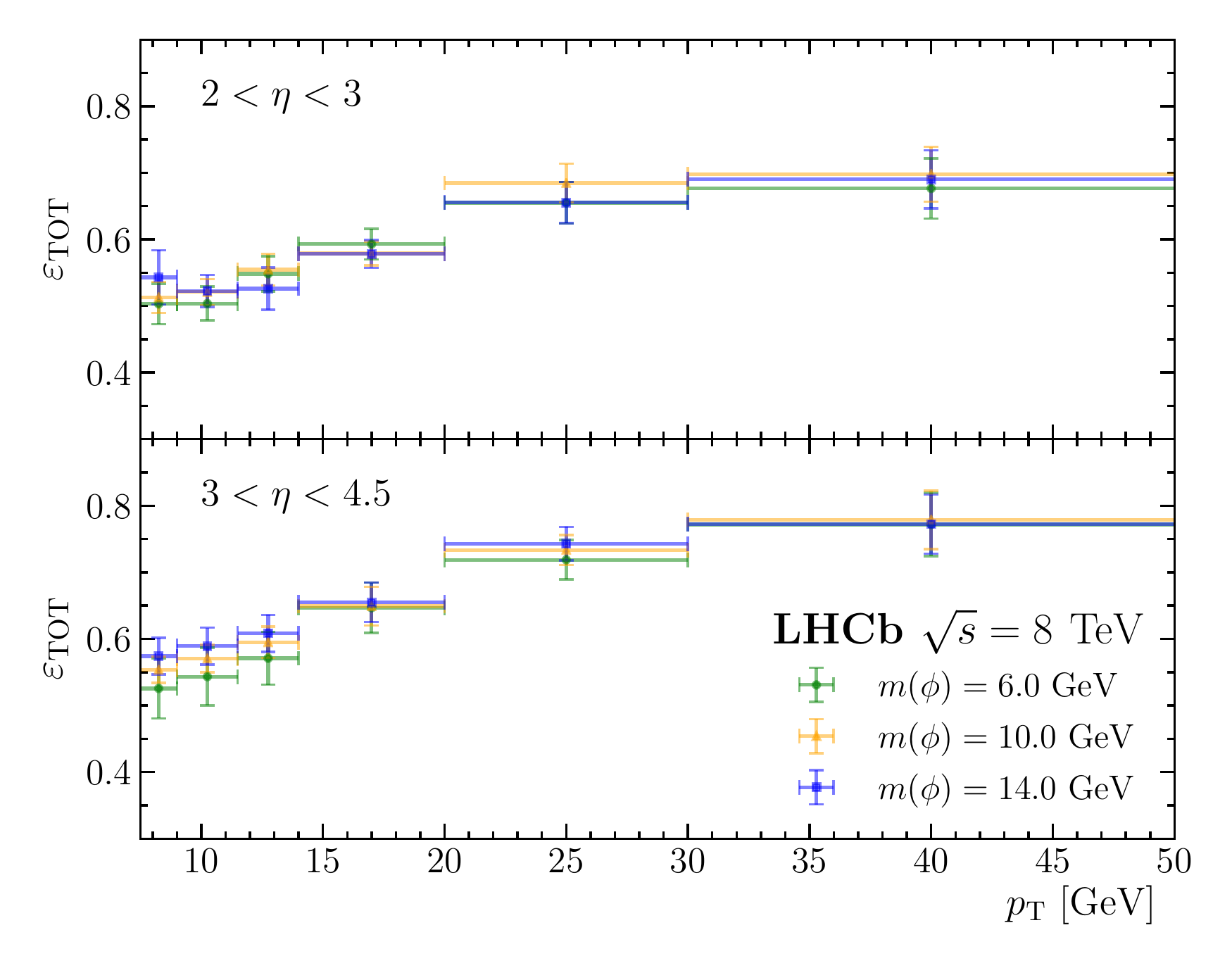}
\caption{Total efficiency as a function of the $\pt$ of the dimuon candidate for the two $\eta(\Azo)$ bins considered, obtained for three different $\Azo$ boson mass hypotheses.}
\label{fig:eff_TOT_flat}
\end{center}
\end{figure}

Finally, the total efficiency in each bin is obtained as the product of the efficiencies described above, where different sources of systematic uncertainties are assumed to be fully correlated.
Efficiencies for the 12 $[\pt, \eta]$ bins and for three different $m(\Azo)$ values are shown in \figref{fig:eff_TOT_flat} for the 8\tev sample. Due to the fiducial region defined, the separation in kinematic bins and the use of the uBoost technique, the efficiencies are minimally correlated with the $\Azo$ boson mass.
A quadratic function is also fitted to the efficiency mass dependence and compared to the mass average. The mean value between the two efficiencies is taken as the nominal value while the difference is assigned as a systematic uncertainty.
The dependence of the efficiency on the $\Azo$ boson kinematics due to the $[\pt,\eta]$ bin size is evaluated by comparing the efficiencies in each bin obtained for $pp\to\Azo$ production to those obtained for a \Azo boson originating from the decay $H\to\Azo\Azo$. The latter production mode gives a vastly different spectrum, with larger \pt and a more central $\eta$ distribution, as shown in \figref{fig:sig_fractions}. The small differences ($1$--$5\%$) in the efficiencies found are assigned as systematic uncertainties. 

For the case where the boson is a vector, a systematic uncertainty of less than 5\% is assigned to account for the dependency of the total efficiency on the boson polarisation.
It is evaluated by weighting the spin-0 $\Azo$ boson sample to match the angular distribution of a vector boson with either longitudinal or transverse polarisation.

\section{Invariant mass fit}
\label{sec:mass}

The $\Azo$ boson signal yield is determined for each mass value with fits to the full dimuon invariant mass spectrum. Due to their complexity, the fits are computed by parallelising the processes on a Graphics Processing Unit (GPU), for which the framework developed in Ref.~\cite{ipanema} is used, where the minimisation is based on Minuit~\cite{James:1975dr}. The natural width of the $\Azo$ boson candidate is assumed to be negligible compared to the detector mass
resolution, which is $\sigma(m_{\mu\mu})/m(\Azo)\approx 0.5\%$ \cite{LHCb-DP-2014-002}. 
These fits are performed simultaneously in the 12 production kinematic bins, sharing some of the parameters. The $\Azo$ boson mass hypotheses are scanned in steps of $\sigma(m_{\mu\mu})/2$. The detector resolution on the dimuon mass is modelled according to $\eta$, $\pt$ and $m(\Azo)$. The resolution model is used to simultaneously fit the \UpsilonNS peaks, which are used for its calibration. Furthermore, in order to increase the invariant mass region scanned and to get as close as possible to
the \UpsilonNS resonances, a precise modelling of the \UpsilonNS mass-distribution tails is needed. For this purpose, the reconstructed dimuon mass, $m_{\mu\mu}$, is modelled by a Gaussian-smeared Hypatia distribution, $\cal S$, which is defined as
\begin{equation}
\begin{split}
  {\cal S}(m_{\mu\mu}, m(\Azo), \sigma_{\rm MS}, \sigma_{\rm SR}, \lambda, \beta, a, n)  = & 
  \frac{1}{\sigma_{\rm MS}} e^{-\frac{1}{2}\left(\frac{m_{\mu\mu}-m(\Azo)}{\sigma_{\rm MS}}\right)^2} \\
  & \otimes {\cal I}(m_{\mu\mu}, m(\Azo), \sigma_{\rm SR}, \lambda, \zeta\rightarrow 0, \beta, a, n)\,,
  \end{split}
\end{equation}

\noindent where $\cal I$ is the Hypatia function \cite{Santos:2013gra}, a generalised Crystal Ball (CB) \cite{Skwarnicki:1986xj} with a hyperbolic core that gives an excellent description of non-Gaussian tails, given by  
\begin{multline}
{\cal{I}}(m_{\mu\mu}, m(\Azo),\sigma_{\rm SR},\lambda,\zeta,\beta,a,n) \propto \\
\begin{cases}
{G(m(\Azo)-a\sigma_{\rm SR})} &\text{ if } \frac{m_{\mu\mu}-m(\Azo)}{\sigma_{\rm SR}} > -a,\\
{G(m(\Azo)-a\sigma_{\rm SR})}{\left(1-m_{\mu\mu}/\left(n\frac{G(m(\Azo)-a\sigma_{\rm SR})}{G'(m(\Azo)-a\sigma_{\rm SR})}-a\sigma_{\rm SR}\right)\right)^{-n}} &\text{ otherwise,}
\end{cases}
\end{multline}

\noindent and $G(x) \equiv G(x, m(\Azo),\sigma_{\rm SR},\lambda,\zeta,\beta)$ is its core, defined as
\begin{multline}
\label{eq:ghypreparam}
G(x; m(\Azo),\sigma_{\rm SR},\lambda,\zeta,\beta) \propto \\
\left((x-m(\Azo))^{2} + A^{2}_{\lambda}(\zeta)\sigma_{\rm SR}^{2}\right)^{\frac{1}{2} \lambda - \frac{1}{4}} e^{\beta (x-m(\Azo))} K_{\lambda - \frac{1}{2}}\left(\zeta \sqrt{1+\left(\frac{x-m(\Azo)}{A_{\lambda}(\zeta)\sigma_{\rm SR}}\right)^{2}}\right),
\end{multline}

\noindent where $G'$ is the derivative of $G$ (defined in Eq.~\ref{eq:ghypreparam}), $K_{\lambda}$ are the cylindrical harmonics or special Bessel functions of the third kind, $\beta$ is the asymmetry of the core, $a$ and $n$ are CB-like radiative-tail parameters, and $A^{2}_{\lambda} = {\zeta K_{\lambda}(\zeta)}/{K_{\lambda+1}(\zeta)}$. The parameter $\zeta$ is known to be small in most cases~\cite{Santos:2013gra}, and thus, is fixed to an arbitrarily small value. In order to reduce the number of free parameters in the simultaneous fits, a parametrisation of the
dependence of the parameters defined above on $\pt$, $\eta$ and $m(\Azo)$ is obtained from the simulation. The parameters $\beta$, $n$
and $a$ are found to be independent of $\pt$, $\eta$ and $m(\Azo)$. The parameter $n$ is fixed to the value obtained from the simulation, while $\beta$ and $a$ are shared among different kinematic bins and mass hypotheses in the fit. Further information about these functions and their parameters can be found in Ref.~\cite{Santos:2013gra}. 

The Gaussian smearing factorises the mass resolution model into two components: the multiple scattering (MS) information, which is encoded in the smearing parameter $\sigma_{\rm MS}$; and the spatial resolution information, which is given by $\sigma_{\rm SR}$ and $\lambda$. In this parametrisation the value of $\sigma_{\rm MS}$ can be fixed from the ramp-up of the mass-error distribution without increasing the dimensionality of the fit. The mass error is obtained in the vertex fit and the ramp-up position is defined as the mass error corresponding to the fifth percentile of the distribution. The parameter $\sigma_{\rm
MS}$ depends on the kinematics, and thus, is modelled in bins of \pt, $\eta$ and $m(\Azo)$ on the continuum background. The $m(\Azo)$ dependence of this MS parameter is studied in bins of dimuon mass and is modelled by a linear fit. The $\sigma_{\rm MS}$ parameter in data is found to be in excellent agreement with the simulation, and therefore, no systematic uncertainty is assigned. 
%%%%

The continuous dimuon background is modelled with an exponential function multiplying Legendre polynomials, $P_k$, up to order $N$. The background shape parameters and yields are fit separately in each $[\pt, \eta]$ bin. For each $m(\Azo)$, the model has to describe the background under the signal peak, $B$, to a precision exceeding its expected statistical fluctuation, $\sqrt{B}$. The background model is tested on a sample composed of simulated Drell-Yan dimuon events and same-sign dimuon data
events. The same-sign dimuon mass spectrum is expected to be representative of the background coming from pions misidentified as muons.
In this mass spectrum, the candidate fit model is required to describe any structure with a width larger than $4\sigma(m_{\mu\mu})$ to a precision better than $0.5\sqrt{B}$. 
Furthermore, a similar test is performed on a large simulated sample of muon pairs coming from heavy-flavour decays. This background component is expected to give narrower structures, therefore the above requirement is reduced to $0.3\sqrt{B}$. 
These requirements are well satisfied by a background model with an exponential function multiplied by Legendre polynomials of order $N=10$, which is taken as reference.

\begin{figure}[th!]
\begin{center}
\includegraphics[width=0.75\textwidth]{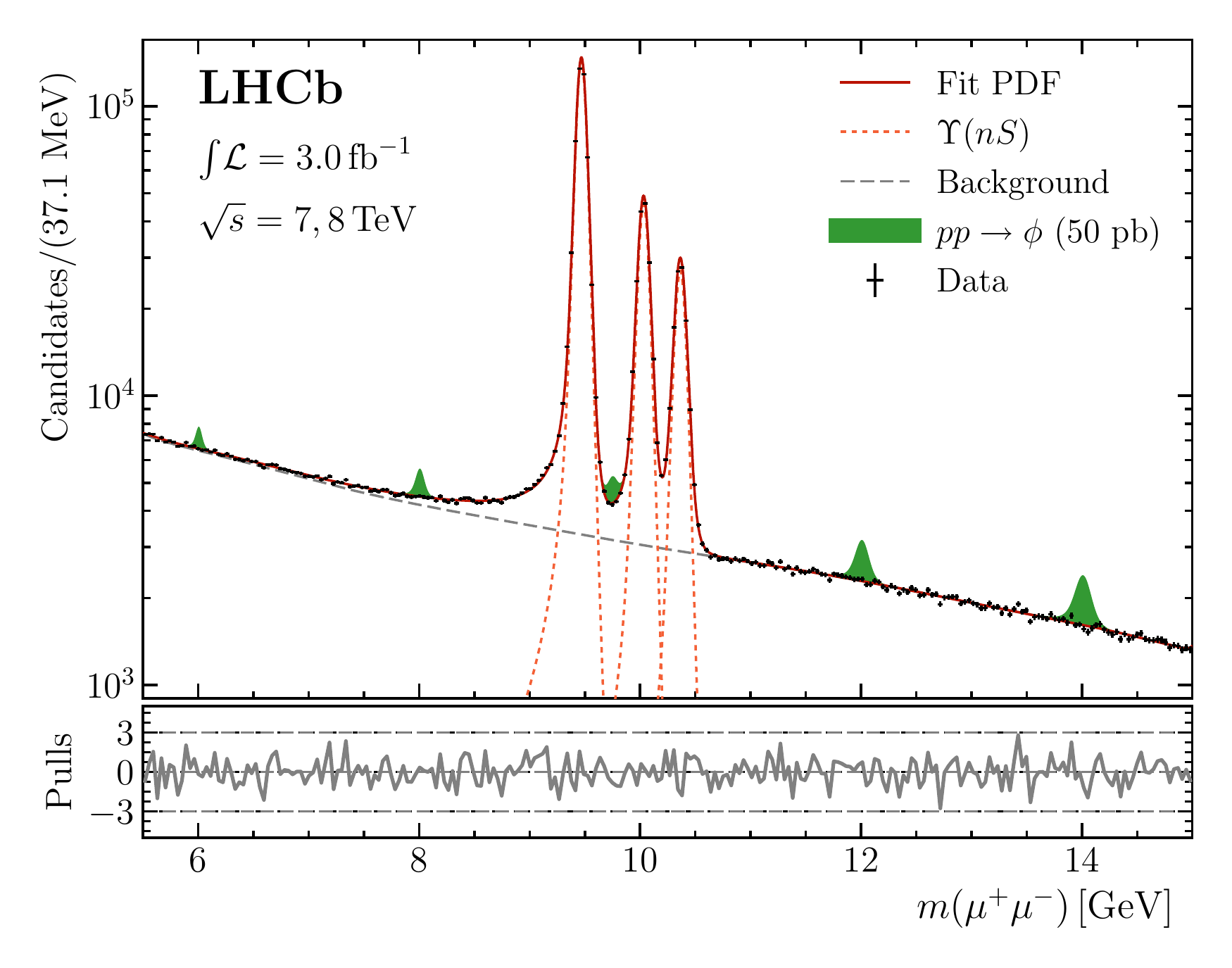}
\includegraphics[width=0.75\textwidth]{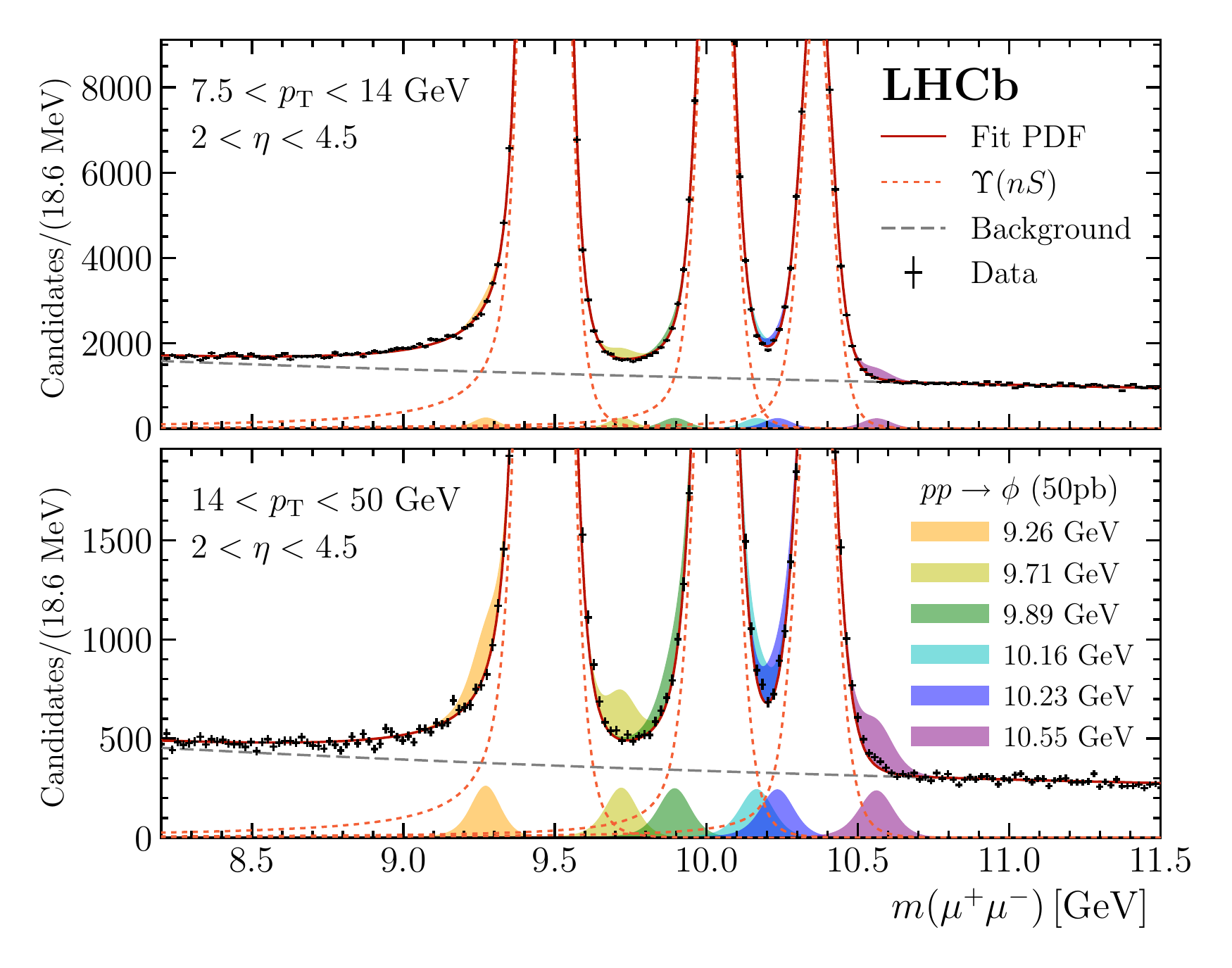}
\caption{(top) Fit to the dimuon invariant mass distribution in the whole scanned region. All $[\pt, \eta]$ bins as well as the 7 and 8~TeV data sets are combined. Peaks for five \Azo boson mass hypotheses are displayed in green, assuming $\sigma(pp \to \Azo) \times \BF( \Azo \to \mup \mun) = 50\pb$. (bottom) A closeup view of the mass spectrum in the \UpsilonNS region together with \Azo boson mass peaks for the tested $m(\Azo)$ values closest to the three \UpsilonNS narrow resonances. To show how the \UpsilonNS mass tails change with the kinematics, two regions of $\pt$ are displayed in the plot.}
\label{fig:mass_fit}
\end{center}
\end{figure}

The results of the fit to data in the whole mass region is shown in \figref{fig:mass_fit}, where all kinematic bins have been combined. The figure also shows how different \Azero boson mass peaks would look like.

The resolution function has 17 free parameters. The fits for $\Azo$ boson mass hypotheses far from the \UpsilonNS peaks are found to be largely independent of the signal model. However, for $m(\Azo)$ close to the \UpsilonNS resonances, the estimate of the background under a possible $\Azo$ boson peak depends on the precise modelling of the \UpsilonNS tails. In particular, significant differences are observed using a resolution function with fewer assumptions on
the kinematic dependence of $\beta$, $a$ and $\lambda$. The $m(\Azo)$ hypotheses for which the two background estimations differ with a significance larger than one standard deviation in any kinematic bin are not considered in the $\Azo$ boson search. 
In addition, any $m(\Azo)$ hypothesis where the fit gives a correlation between the signal yield and any of the \UpsilonNS yields in excess of $20\%$, is also excluded from the search.  

\section{Results}
\label{sec:results}

The fit results are found to be compatible with the background-only hypothesis. Upper limits at $95\%$ Confidence Level (CL) are set on spin-0 $\Azo$ bosons produced directly from the $pp$ collision. Pseudoexperiments are generated based on the fitted background probability distribution functions and upper limits are
determined using the CLs approach~\cite{CLs1, CLs2}. Measured integrated luminosities, simulated signal production spectra and the model-independent efficiencies given in Sec.~\ref{sec:efficiencies} are used to compute expected signal yields in each $[\pt, \eta]$ bin. Systematic and statistical uncertainties on the efficiency are propagated to the limit calculation, summing them in quadrature and taking into account their correlations among different bins. The integrated luminosities for the 7 and 8\tev samples are measured~\cite{LHCB-PAPER-2014-047} with a precision of $1.7\%$ and $1.2\%$, respectively.

\begin{figure}[t]
\begin{center}
\includegraphics[width=0.8\textwidth]{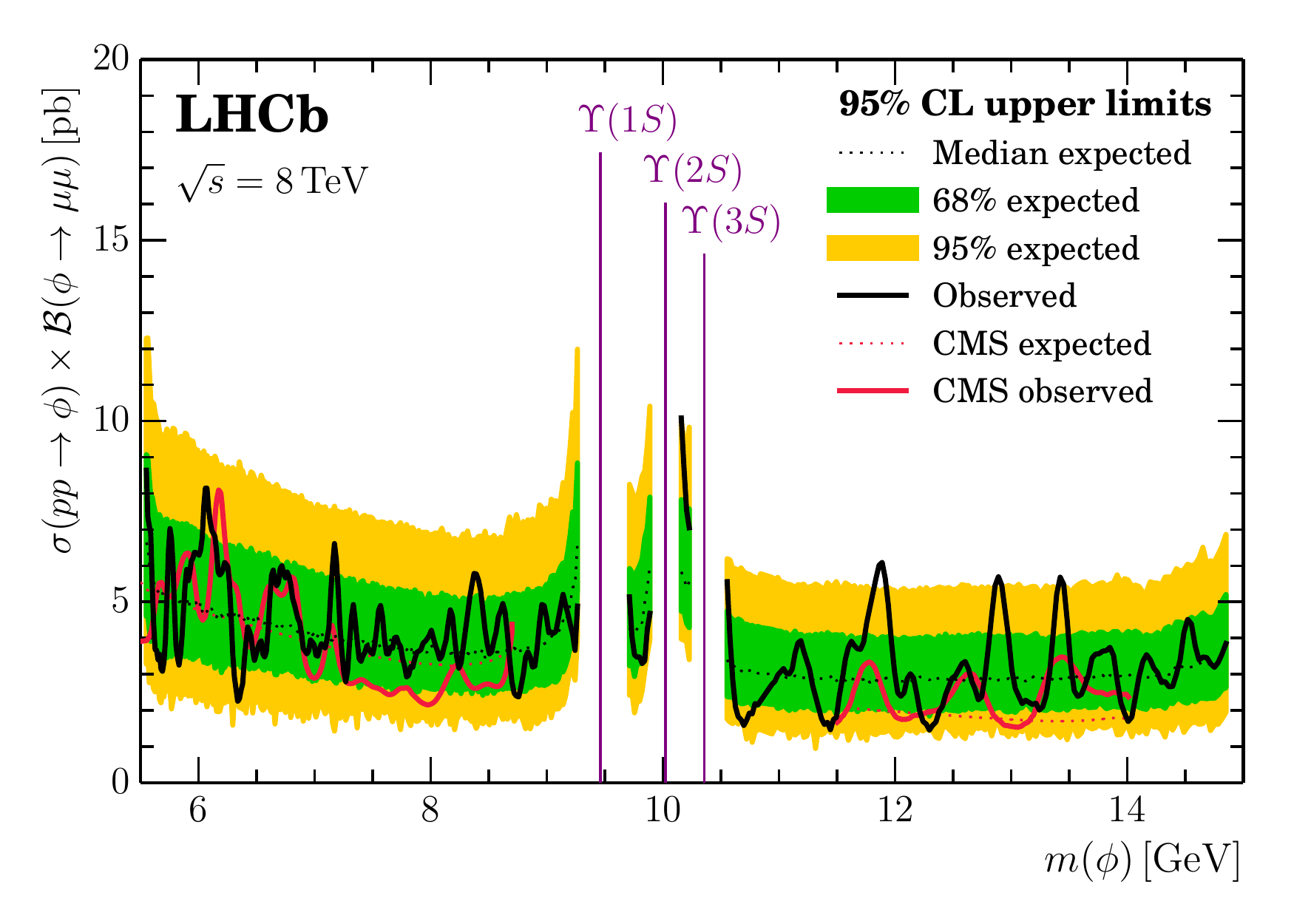}
\caption{Upper limits on the direct production of a spin-0 boson decaying to $\mu^+\mu^-$ in 8~TeV $pp$ collisions. }
\label{fig:limits}
\end{center}
\end{figure}

The production kinematics for spin-0 \Azo bosons are simulated using the MSSM pseudo-scalar production implemented in \pythia~8~\cite{Sjostrand:2007gs}. Gluon-gluon fusion dominates, contributing more than $90\%$ to the production cross-section in the whole $\Azo$ boson mass range. 
In order to set limits on new spin-0 particles in terms of couplings, interference effects with spin-0 bottomonium states should be considered~\cite{Haisch:2016hzu}, but this is beyond the scope of this analysis. Therefore, upper limits are set on 
the product of the production cross-section and the dimuon branching fraction, $\sigma(pp\to\Azo) \times \BR{\Azo\to\mumu}$.
Since the cross-section depends on the collision energy, the limits are set for $\sqrt{s}=8\tev$ and the result from $7\tev$ is combined by taking the expected fraction of cross-sections as a function of $m(\Azo)$, based on the framework detailed in Ref.~\cite{Haisch:2016hzu}. This ratio of cross-sections is roughly equal to the ratio of collision energies and has a small dependence on $m(\Azo)$ of order $4\%$ within the mass range considered. The observed limits are given in~\figref{fig:limits} along with the range of limits expected for the background-only hypothesis.

In \appref{app:interpretation} the upper limits are interpreted for \Azo bosons coming from the decay of the 125\gev Higgs boson to two \Azo bosons and for vector $A^\prime$ bosons with Drell-Yan production. If the vector $A^\prime$ boson is interpreted as a dark photon, these are the first limits in the region between 9.1 and 10.6 \gev.
Furthermore, reinterpretation of the limits in any other model involving the production of a dimuon resonance in the mass range considered is possible by using the information given in the supplemental material. 

\section{Conclusions}
\label{sec:conclusions}

In summary, a search is presented for a hypothetical light dimuon resonance, produced in $pp$ collisions recorded by the LHCb detector at centre-of-mass energies of 7 and 8\tev. A sample of dimuon candidates with invariant mass between 5.5 and 15\gev corresponding to an integrated luminosity of 3.0\invfb is used. 
No evidence for a signal is observed and limits are placed on a benchmark model involving a new light spin-0 boson, \Azo, decaying to a pair of muons. For the case in which the \Azo boson is produced directly in the $pp$ collision, the limits obtained are comparable with the best existing. Furthermore, by exploiting the excellent LHCb dimuon mass resolution and a detailed study of the \UpsilonNS mass tails, limits are set in a previously unexplored range of
$m(\Azo)$ between 8.7 and 11.5\gev. This search is designed to be largely model independent and tools are given in the supplemental material that allow for a simple reinterpretation of the result for different models. 
These results showcase the sensitivity of the LHCb experiment to light spin-0 bosons produced in $pp$ collisions and its capability of closing the gaps in the invariant mass distributions by means of a superior mass resolution.

% Comment this in for paper darfts; do not include this in analysis note and conference reports
\section*{Acknowledgements}
%
% These Acknowledgements valid from 20-Mar-2018
%
\noindent We express our gratitude to our colleagues in the CERN
accelerator departments for the excellent performance of the LHC. We
thank the technical and administrative staff at the LHCb
institutes. We acknowledge support from CERN and from the national
agencies: CAPES, CNPq, FAPERJ and FINEP (Brazil); MOST and NSFC
(China); CNRS/IN2P3 (France); BMBF, DFG and MPG (Germany); INFN
(Italy); NWO (The Netherlands); MNiSW and NCN (Poland); MEN/IFA
(Romania); MinES and FASO (Russia); MinECo (Spain); SNSF and SER
(Switzerland); NASU (Ukraine); STFC (United Kingdom); NSF (USA).  We
acknowledge the computing resources that are provided by CERN, IN2P3
(France), KIT and DESY (Germany), INFN (Italy), SURF (The
Netherlands), PIC (Spain), GridPP (United Kingdom), RRCKI and Yandex
LLC (Russia), CSCS (Switzerland), IFIN-HH (Romania), CBPF (Brazil),
PL-GRID (Poland) and OSC (USA). We are indebted to the communities
behind the multiple open-source software packages on which we depend.
Individual groups or members have received support from AvH Foundation
(Germany), EPLANET, Marie Sk\l{}odowska-Curie Actions and ERC
(European Union), ANR, Labex P2IO and OCEVU, and R\'{e}gion
Auvergne-Rh\^{o}ne-Alpes (France), Key Research Program of Frontier
Sciences of CAS, CAS PIFI, and the Thousand Talents Program (China),
RFBR, RSF and Yandex LLC (Russia), GVA, XuntaGal and GENCAT (Spain),
Herchel Smith Fund, the Royal Society, the English-Speaking Union and
the Leverhulme Trust (United Kingdom).

\clearpage

{\noindent\normalfont\bfseries\Large Appendices}

\appendix

\section{Results for other models}
\label{app:interpretation}

\begin{figure}[h]
\begin{center}
\includegraphics[width=0.49\textwidth]{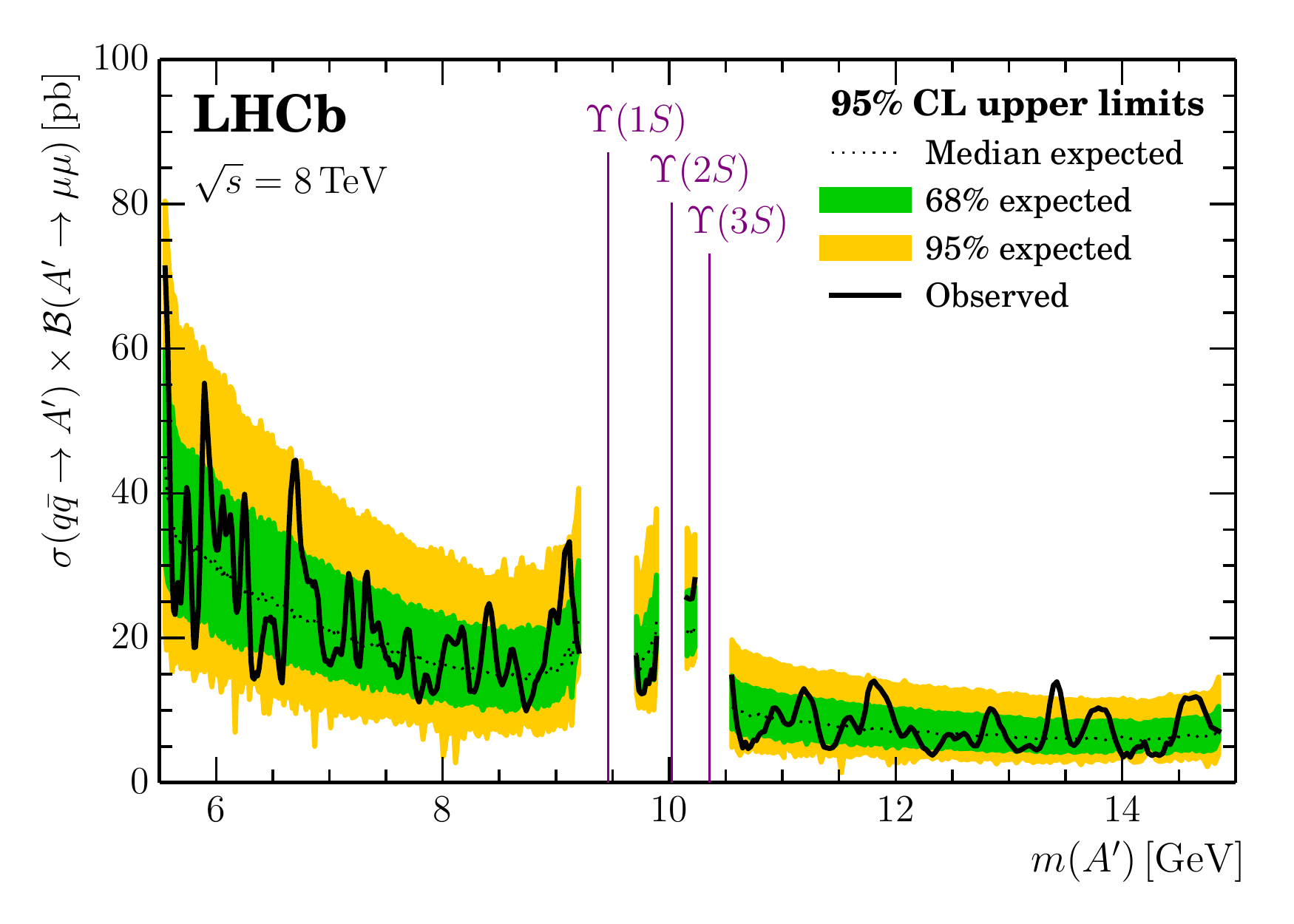}
\includegraphics[width=0.49\textwidth]{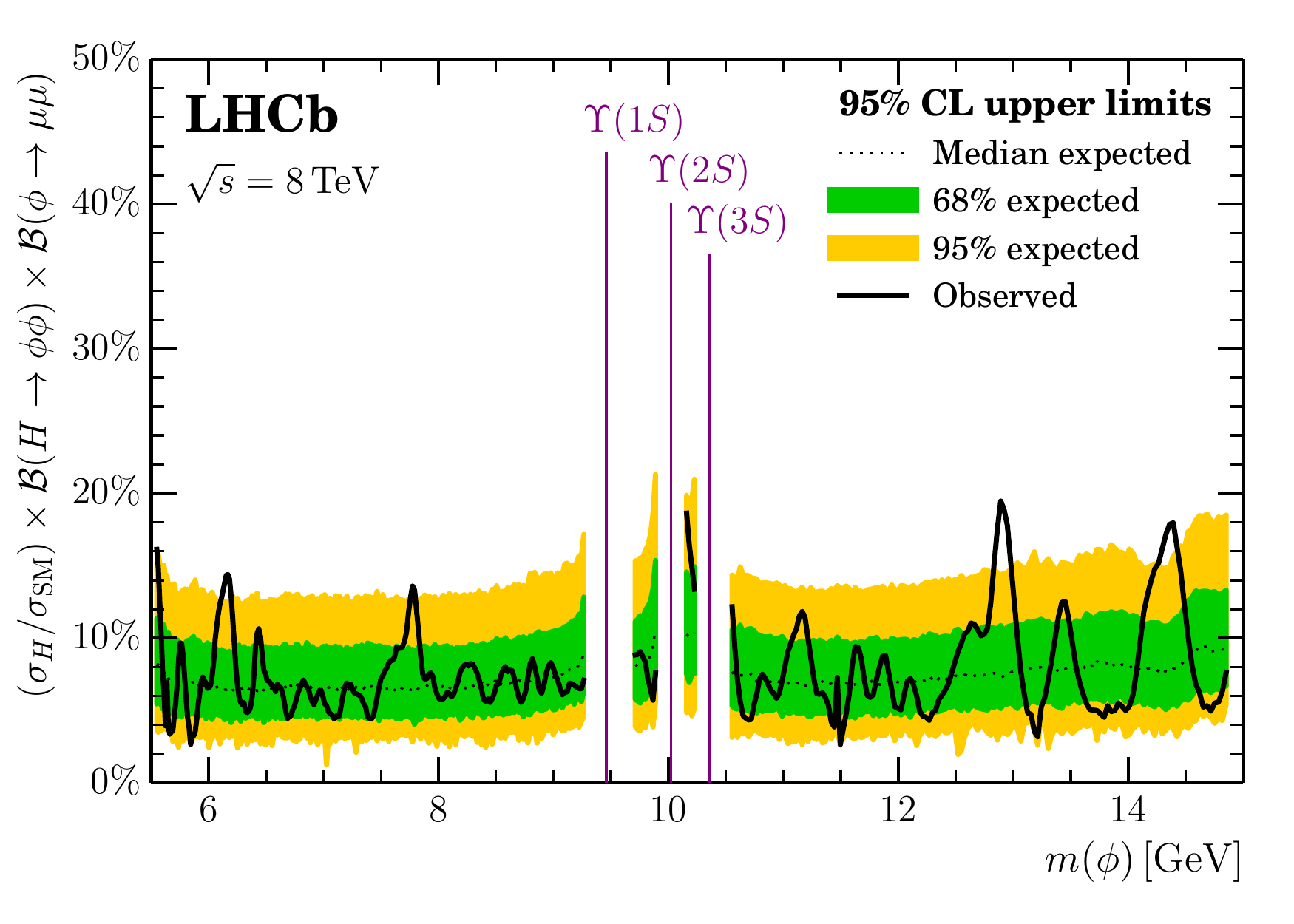}
\caption{(left) Upper limits on the production of vector $A^\prime$ bosons produced in 8~TeV $pp$-collisions through Drell-Yan and decaying to $\mu^+\mu^-$. (right) Upper limits on the branching fraction of a SM Higgs decaying to two \Azo bosons followed by the decay of one of the two to  $\mu^+\mu^-$.}
\label{fig:otherlimits}
\end{center}
\end{figure}

Two additional boson production models are tested and the resulting upper limits are shown in \figref{fig:otherlimits}. The first model is a vector boson, $A^\prime$, produced via Drell-Yan $q\bar{q}\to A^\prime$ and decaying to a pair of muons. The Drell-Yan production kinematics are taken from \pythia~8~\cite{Sjostrand:2007gs}. These results can be interpreted as limits on dark photons since their production mode is expected to be dominated by Drell-Yan in this region of masses.

In the second model the signal is assumed to come from the decay of the 125\gev Higgs to two spin-0 \Azo bosons. 
Only one of the two \Azo bosons is required to decay to a dimuon final state, so the limit is set on ${(\sigma_H/\sigma_{\rm SM})\times\BR{H\to\Azo\Azo}\times\BR{\Azo\to\mumu}}$, where $\sigma_H$ is the 125\gev Higgs cross-section and $\sigma_{\rm SM}$ is its value as computed in the SM. The combination of 7 and 8\tev results is obtained by taking for $\sigma_{\rm SM}$ the SM gluon-gluon fusion cross-sections for a 125\gev Higgs from Ref.~\cite{Heinemeyer:2013tqa} and assuming that
$\sigma_H/\sigma_{\rm SM}$ is independent on the centre-of-mass energy $\sqrt{s}$. 

The most significant excess is $2.9~\sigma$ at $m(\Azo)\simeq 12.92\gev$ in the $H\to\Azo\Azo$ production model hypothesis and has a $p$-value of $14\%$, after accounting for the trials factor due to the large mass range tested in comparison to the dimuon mass resolution. 

\clearpage

\addcontentsline{toc}{section}{References}
\setboolean{inbibliography}{true}
\bibliographystyle{LHCb}
\bibliography{main,LHCb-PAPER,LHCb-CONF,LHCb-DP,LHCb-TDR}

\newpage

\newpage
\centerline{\large\bf LHCb collaboration}
\begin{flushleft}
\small
R.~Aaij$^{43}$,
B.~Adeva$^{39}$,
M.~Adinolfi$^{48}$,
Z.~Ajaltouni$^{5}$,
S.~Akar$^{59}$,
P.~Albicocco$^{19}$,
J.~Albrecht$^{10}$,
F.~Alessio$^{40}$,
M.~Alexander$^{53}$,
A.~Alfonso~Albero$^{38}$
% ...

{\footnotesize \it
% ...
$ ^{5}$Clermont Universit{\'e}, Universit{\'e} Blaise Pascal, CNRS/IN2P3, LPC, Clermont-Ferrand, France\\
% ...
$ ^{10}$Fakult{\"a}t Physik, Technische Universit{\"a}t Dortmund, Dortmund, Germany\\
% ...
$ ^{19}$Laboratori Nazionali dell'INFN di Frascati, Frascati, Italy\\
% ...
$ ^{38}$ICCUB, Universitat de Barcelona, Barcelona, Spain\\
$ ^{39}$Instituto Galego de F{\'\i}sica de Altas Enerx{\'\i}as (IGFAE), Universidade de Santiago de Compostela, Santiago de Compostela, Spain\\
$ ^{40}$European Organization for Nuclear Research (CERN), Geneva, Switzerland\\
% ...
$ ^{43}$Nikhef National Institute for Subatomic Physics, Amsterdam, The Netherlands\\
% ...
$ ^{48}$H.H. Wills Physics Laboratory, University of Bristol, Bristol, United Kingdom\\
% ...
$ ^{53}$School of Physics and Astronomy, University of Glasgow, Glasgow, United Kingdom\\
% ...
$ ^{59}$University of Cincinnati, Cincinnati, OH, United States\\
% ...
}
\end{flushleft}

\end{document}